\documentclass[11pt]{article}
\usepackage{graphicx}

\def\RDz{R_{D}}
\def\NDtr{N^D_{\rm tr}}

\def\Dmiss{D}

\def\P{{\cal P}}
\def\F{{\cal F}}
\def\M{{\cal M}}
\def\R{{\cal R}}
\def\E{{\cal E}}
\def\T{{\cal T}}

\def\comb{}

\def\NDtr{N^D_{\rm tr}}
\def\dz{\Delta z}

\def\dt{\Delta t}
\def\Dm{\Delta m}
\def\Dt{\Delta t}

\def\tauB{\tau_{_{\hspace{-2pt}B^0}}}

\def\A{{A}}

\def\peak{{{\rm peak}\Bz}}
\def\Dstst{{D^{**}}}
\def\bdstrho{\Bzb \rightarrow {\Dstar} ^+ \rho^-}
\def\fisher{F}
\def\mrho{m(\rho)}
\def\mmiss{m_{\rm miss}}
\def\rhoHelic{\cos\theta_{\rho}}
\def\dstHelic{\cos\theta_{\Dstar}}
\def\aOne{{\Bzb \rightarrow {\Dstar} ^+ a_1^-}}
\def\DstAOne{{\Dstar}a_1}
\def\BDstst{\overline B\rightarrow D^{**} \rho^-}
\def\zrec{z_{\rm rec}}
\def\ztag{z_{\rm other}}

\def\Dt{\Delta t}

\def\DtErr{\sigma_{_{\Dt}}}

\def\mc{Monte Carlo}
\def\lumi{20.7}
\def\lumioff{2.6}
\def\nB{22.7 million}

\newcommand{\BABARPubYear}    {02}

\newcommand{\BABARConfNumber} {003}
\newcommand{\SLACPubNumber} {9169}

\input babarsym
\long\def\inst#1{\par\nobreak\kern 4pt\nobreak
    {\it #1}\par\vskip 10pt plus 3pt minus 3pt}

\begin{document}
{\pagestyle{empty}

\begin{flushright}
SLAC-PUB-\SLACPubNumber \\
\babar-CONF-\BABARPubYear/\BABARConfNumber \\
 \end{flushright}

\par\vskip 5cm

\begin{center}
\Large \bf Measurement of the \boldmath $\Bz$ Lifetime 
	with Partial Reconstruction of 
	$\bdstrho$
\end{center}
\bigskip

\begin{center}
\large The \babar\ Collaboration\\
\mbox{ }\\
\today
\end{center}
\bigskip \bigskip

\begin{center}
\large \bf Abstract
\end{center}
A sample of about 5500 $\bdstrho$ and 700 $\aOne$ events is
identified, using the technique of partial reconstruction, among \nB\
$\BB$ pairs collected by the \babar\ experiment at the \pep2\ storage
ring. With these events, the $\Bz$ lifetime is measured to be $1.616
\pm 0.064 \pm 0.075$~ps. This measurement serves as validation for the 
procedures required to measure
$\sin(2\beta + \gamma)$ with partial reconstruction of $\bdstrho$.

\vfill
\begin{center}
Presented at the XXXVII$^{th}$ Rencontres de Moriond on 
QCD and Hadronic Interactions,\\ 
3-16---3/23/2002, Les Arcs, Savoie, France 
\end{center}

\vspace{1.0cm}
\begin{center}
{\em Stanford Linear Accelerator Center, Stanford University, 
Stanford, CA 94309} \\ \vspace{0.1cm}\hrule\vspace{0.1cm}
Work supported in part by Department of Energy contract DE-AC03-76SF00515.
\end{center}
}

\newpage

\begin{center}
\small

The \babar\ Collaboration,
\bigskip

B.~Aubert,
D.~Boutigny,
J.-M.~Gaillard,
A.~Hicheur,
Y.~Karyotakis,
J.~P.~Lees,
P.~Robbe,
V.~Tisserand,
A.~Zghiche
\inst{Laboratoire de Physique des Particules, F-74941 Annecy-le-Vieux, France }
A.~Palano,
A.~Pompili
\inst{Universit\`a di Bari, Dipartimento di Fisica and INFN, I-70126 Bari, Italy }
% ...

\end{center}\newpage


\section{Introduction}
The neutral $B$ meson decay modes $B \rightarrow {\Dstar}^+ h^-$, where
$h^-$ is a light hadron ($\pi^-, \rho^-, a_1^-$), have been
proposed~\cite{ref:book} for use in theoretically clean measurements of the
Cabibbo-Kobayashi-Maskawa~\cite{ref:km} unitarity triangle parameter
$\sin(2\beta+\gamma)$.
Since the time-dependent CP asymmetries in these modes are expected to
be of order 2\%, large data samples and multiple decay channels are
required for a statistically significant measurement. The low
efficiency of reconstructing the $\Dz$ produced in the ${\Dstar}$
decay leads to significant loss of signal events. Partial reconstruction of the
$B$ meson results in substantially larger efficiency, albeit with
higher backgrounds. The overall $\sin(2\beta+\gamma)$ sensitivity is
expected to be roughly similar for partial and full reconstruction,
and about 90\% of the partially reconstructed events cannot
be fully reconstructed. Therefore, both full and partial
reconstruction can and should be used for the $\sin(2\beta+\gamma)$
measurement.

The measurement of the $\Bz$ lifetime, presented here, constitutes a
first step toward measuring $\sin(2\beta+\gamma)$ with partial
reconstruction of $\bdstrho$. In this analysis, we have developed the
procedures for candidate reconstruction, background characterization,
vertexing, and fitting, all critical components of the time-dependent
$\sin(2\beta+\gamma)$ analysis.

\section{The \babar\ Detector and Data Sample}
The data used in this analysis were collected with the \babar\
detector at the \pep2\ storage ring. The data consist of \nB\ $\BB$
pairs, corresponding to an integrated luminosity of \lumi~$\invfb$
recorded at the $\FourS$ resonance.
In addition, \lumioff~$\invfb$ were collected about
40~MeV below the resonance. This off-resonance sample is used to study
the continuum, $\epem \rightarrow \qqbar$ background, where $q = \{u,d,s,c\}$.

The \babar\ detector, described in detail elsewhere~\cite{ref:nim},
consists of five sub-detectors.  Surrounding the beam-pipe is a
five-layer silicon vertex tracker (SVT), providing precision measurements
of the positions of charged particles close to the interaction point
and tracking of charged particles with low transverse momentum. 
Outside the support tube that surrounds the SVT is a 40-layer drift
chamber (DCH), filled with an 80:20 helium-isobutane gas mixture. The
DCH provides charged particle momentum measurements in a 1.5~T
magnetic field, and ionization energy loss measurements that
contribute to charged particle identification. Surrounding the DCH is
a detector of internally reflected Cherenkov light (DIRC), providing
charged particle identification. Outside the DIRC is a CsI(Tl)
electromagnetic calorimeter (EMC), used mainly to detect and measure
the energy of photons and to provide electron identification. The EMC
is surrounded by a superconducting coil, which generates the magnetic
field for tracking. Outside the coil, the flux return iron is
instrumented with resistive plate chambers, used mainly for muon
identification.

\pep2\ is an asymmetric energy storage ring, with positron and
electron beam energies of about 3.11 and 9.0~GeV. The center-of-mass
(CM) frame of the average $\epem$ collision is therefore boosted
along the $z$ direction in the lab frame, enabling time-dependent 
measurements involving reconstruction of $B$ mesons.

\section{Analysis Method}
\subsection{Partial Reconstruction}

To partially reconstruct a $\bdstrho$ candidate\footnote{Charge
conjugate decays are also implied.} , only the $\rho$ and the $\pi_s$,
the soft pion from the decay ${\Dstar}^+ \rightarrow D^0\pi_s^+$
decay, are reconstructed. The angle between the momenta of the $B$ and
the $\rho$ in the CM frame is then computed:
\begin{equation}
\cos\theta_{B\rho} = 
	{M_{{\Dstar}^+}^2 - M_{\Bz}^2 - M_{\rho}^2 + E_{\rm CM} E_{\rho}
			\over 
	 2 P_B |\vec p_{\rho}|
	},
\label{eq:cosTheta}
\end{equation}
where $M_x$ is the mass of particle~$x$, $E_\rho$ and $\vec p_\rho$
are the measured CM energy and momentum of the $\rho$, $E_{\rm CM}$ is
the total CM energy of the beams, and 
$P_B = \sqrt{E_{\rm CM}^2/4 - M_{\Bz}^2}$.
Given $\cos\theta_{B\rho}$ and the measured four-momenta of the
$\pi_s$ and the $\rho$, the $B$ four-momentum may be calculated up to
an unknown azimuthal angle $\phi$ around ${\vec p}_{\rho}$. For every
a value of $\phi$, the expected $\Dmiss$ four-momentum, ${\cal
P}_D(\phi)$, is determined from four-momentum conservation, and the
$\phi$-dependent ``missing mass'' is calculated,
$m(\phi) \equiv \sqrt{|{\cal P}_D(\phi)|^2}$.
With $m_{\rm max}$ and $m_{\rm min}$ being the maximum and minimum values
of $m(\phi)$ obtained by varying $\phi$,
we define the missing mass, 
$\mmiss \equiv {1 \over 2}\left[m_{\rm max} + m_{\rm min}\right]$.
This variable peaks around $M_{\Dz}$, the nominal $\Dz$ mass, for
signal events, with a spread of about 3.5~MeV, while background events
are more broadly distributed.

We define the $\Dstar$ helicity angle $\theta_{\Dstar}$ to be the
angle between the directions of the $D$ and the $B$ in the $\Dstar$
rest frame. The value of $\dstHelic$ is computed as in
Ref.~\cite{ref:cleo-dstpi}.
The $\rho$ helicity angle $\theta_{\rho}$ is defined as the angle
between the directions of the $\piz$ (from the decay of the $\rho$)
and the CM frame in the $\rho$ rest frame.
Since the $\bdstrho$ decay is $(87.8 \pm 4.5)\%$ longitudinally
polarized~\cite{ref:polar}, the $\rhoHelic$ and $\dstHelic$
distributions of signal events peak toward $\pm 1$. Background
events have fairly flat distributions, with $\BB$ background
increasing in the region of soft $\pi_s$ and $\piz$,
making $\dstHelic$ and $\rhoHelic$ useful for background suppression.
With knowledge of $\dstHelic$ and $\cos\theta_{B\rho}$, the
3-momentum of the $D$ is determined, up to a two-fold ambiguity and detector
resolution effects.

\subsection{Event Selection}

In order to suppress the continuum background, we select events in which the
ratio of the 2nd to the 0th Fox-Wolfram coefficients~\cite{ref:R2} is
smaller than 0.35.
Charged $\rho$ candidates are identified by their decay to a ``hard''
charged pion, $\pi_h$, and a $\piz$. To suppress fake $\piz$
candidates, the $\piz$ momentum in the CM frame is required to be
greater than 400~MeV. The invariant mass of the $\piz$ candidate must
be within 20~MeV of the nominal $\piz$ mass.
Both the $\pi_h$ and $\pi_s$ candidate tracks are required to originate
within 1.5~cm of the interaction point in the $x-y$ plane (the
plane perpendicular to the beam) and within $\pm 10$~cm
of the interaction point along the direction of the beams.
Tracks are rejected if their specific ionization
and/or Cherenkov angle indicate that they are highly likely to be a
kaon or a lepton.
The invariant mass $\mrho$ of the $\rho$  candidate must be
between 0.45 and 1.1~GeV.

The computed cosine of the angle between the $B$ and $\rho$ must satisfy
$|\cos\theta_{B\rho}| \le 1$. 
To suppress combinatoric background, we require $|\rhoHelic| > 0.3$
and $|\dstHelic| > 0.3$, and reject events in the range $\rhoHelic >
0.3$ and $\dstHelic < -0.3$.

For each $\rho^\pm \pi_s^\mp$ pair that satisfies the above requirements,
$\mmiss$ is calculated. The pair with the smallest value of
$|\mmiss-M_{\Dz}|$ in the event is selected, and all others are
discarded.
Similarly, the ``wrong-sign'' $\rho^\pm \pi_s^\pm$ pair with the smallest 
$|\mmiss-M_{\Dz}|$ is retained for background studies, as described below.
Right-sign events in the range $1.810 < \mmiss < 1.840$~GeV
are classified as sideband events, and are used for background
studies. 
Right- and wrong-sign events satisfying $\mmiss > 1.845$~GeV are
classified as signal region events. All other events are discarded.
The efficiency of signal events to satisfy all the signal region criteria is 6.4\%.

Events that could be fully reconstructed in the $\Dz$ decay modes
$\Dz \rightarrow K^- \pi^+$ or $K^- \pi^+ \piz$ are tagged as fully
reconstructed. In addition to satisfying the partial reconstruction
criteria above, fully reconstructed events are identified by
requiring that
the reconstructed $D$ invariant mass be within 40~MeV of
$M_{\Dz}$,
the difference between the $\Dstar$ and $D$ invariant masses be
between 142~and 150~MeV,
the reconstructed CM $B$ energy be within 50~MeV of $E_{\rm CM}/2$, 
and $\mes > 5.25$~GeV, where $\mes \equiv \sqrt{E_{\rm CM}^2/4 - p_B^2}$
is the beam energy substituted mass and $p_B$ is the reconstructed CM
momentum of the $B$ meson.

\subsection{Measurement of the Decay Time Difference}
The decay position $\zrec$ of the partially reconstructed $B$
candidate along the beam direction is determined by constraining the
$\pi_h$ track to originate from the beam-spot in the $x-y$ plane. To
account for the $B$ meson flight in the $x-y$ plane, $30~\mu {\rm m}$
are added in quadrature to the beam-spot size. The $\pi_s$ is not 
used in this vertex fit in order to simplify the classification of 
the background, and since its contribution to the vertex precision
is small, due to multiple scattering.

The decay position $\ztag$ of the other $B$ meson along the beam
direction, is obtained with all tracks excluding the $\pi_h$,
the $\pi_s$, and any track whose CM angle with respect to either
of the two calculated directions of the $D$ is smaller than
1~radian. This ``cone cut'' reduces $\NDtr$, the number of $D$
daughter tracks used in the other $B$ vertex. The remaining tracks are
fit with a constraint to the beam-spot in the $x-y$ plane.  The track
with the largest contribution to the $\chi^2$ of the vertex, if
greater than 6, is removed from the vertex, and the fit is carried out
again, until no track fails this requirement.

The decay time difference $\dt = (\zrec - \ztag) / \gamma\beta c$ is
then calculated, where $\gamma\beta$ is the CM frame
boost. The value of $\gamma\beta$ is continuously determined from the beam
energies, and averages 0.55. 
The estimated error $\DtErr$ in the measurement of $\Dt$ is
calculated from the parameters of the tracks used in the two vertex
fits.

Events are rejected if the $\chi^2$ probability of the $\pi_h$ vertex
fit is smaller than 1\%, or if the $\chi^2$ probability of the other
$B$ vertex is smaller than 0.5\%. We also require $|\Dt| < 15$~ps and
$0.3 < \DtErr < 4$~ps.

The quantities $\zrec$ and $\ztag$ are computed in the same way for fully
reconstructed candidates as they are for partially reconstructed
candidates. Using the \mc\ simulation, it is verified that with the
1~radian cone cut, the $\NDtr$ distribution of events that are fully
reconstructed in the mode $\Dz \rightarrow K^- \pi^+$ or $K^- \pi^+
\piz$ is in good agreement with the distribution of partially
reconstructed events.

\subsection{Backgrounds}
The types of events in the partially reconstructed on-resonance sample
are classified as follows:
\begin{enumerate}
\item Signal: $\bdstrho$ events, in which the $\pi_h$ is correctly
identified. This requirement ensures that $\zrec$ is the decay
position of the $B$ meson, up to the effect of detector resolution.
The $\pi_s$ or the $\piz$ candidates may be mis-reconstructed.
\item $\aOne$: $\aOne$ events, in which the $\pi_h$ is a daughter of
the $a_1$, and hence originates from the decay point of the $B$ meson.
\item Peaking $\BzBzb$ background: $\bdstrho$ and
some $\aOne$ events, in which the $\pi_h$ originates from the other
$B$ meson, resulting in the measurement $\dt=0$, up to the effect of
detector resolution and the selection of tracks used in the other $B$
vertex. The $\mmiss$ distribution of these events peaks around $M_{\Dz}$, 
similar to signal events.
\item ``Combinatoric'' $\BB$ background: Random combinations
	of $\pi_h$, $\piz$, and $\pi_s$ candidates, possibly including true
	$\rho$ decays.
\item $\BDstst$, where $\Dstst$ stands for a charged or neutral 
resonance with mass in the range $2.4 - 2.5$~GeV decaying into ${\Dstar}^+\pi$.
\item Continuum, $\epem \rightarrow \qqbar$ events.
\end{enumerate}

\subsection{Probability Density Function}
An unbinned maximum likelihood fit is used to obtain the $\Bz$
lifetime $\tauB$ from the data. The fit is performed simultaneously to
on- and off-resonance data, and to on-resonance events which were
fully reconstructed. The probability density function (PDF) is a
function of $\dt$, $\DtErr$, and four ``kinematic'' variables:
1)~$\mmiss$; 2)~$\mrho$; 3) $\mes$; and 4) $\fisher$, a Fisher
discriminant that helps distinguish between $\BB$ and $\qqbar$
events. The value of $\fisher$ is computed from the total CM energy
flow of tracks and neutral EMC clusters (excluding the $\rho$ and the
$\pi_s$) into nine volumes, defined by nine $10^\circ$-wide concentric
cones centered around the $\rho$ CM momentum $\vec p_h$. Each cone is
folded to combine the energy flow in both hemispheres with respect to
$\vec p_h$.

The PDF of partially reconstructed on-resonance events is a sum of
terms corresponding to the different event types:
\begin{eqnarray}
\P(\vec \xi) &=& 
    f_{\rm signal}\, \P_{\rm signal} (\vec \xi) 
+ f_{\DstAOne}\, \P_{\DstAOne} (\vec \xi) \nonumber\\
&+& f_{\peak}\, \P_{\peak} (\vec \xi) 
+ f_{{\rm \comb}\BB}\, \P_{{\rm \comb}\BB} (\vec \xi) \nonumber\\
&+& f_{\Dstst}\, \P_{\Dstst} (\vec \xi) 
+ f_{\qqbar}\, \P_{\qqbar} (\vec \xi),
\label{pdf:inc}
\end{eqnarray}
where $\vec \xi \equiv (\mmiss, \mrho, \fisher, \Dt, \DtErr)$
is the vector of fit variables,
\begin{equation}
\P_i(\vec \xi) \equiv 
\M_i(\mmiss) \;
\R_i(\mrho)  \;
\F_i(\fisher)  \;
\T_i(\Dt, \DtErr) 
\end{equation}
is the PDF corresponding to event type $i$, 
and $f_i$ is the fraction of events of type $i$ in the data sample, where 
$\sum_i f_i = 1$.

The PDF of the off-resonance sample is $\P_{\qqbar}$, which is also
used to describe the continuum component of the on-resonance events in
Eq.~(\ref{pdf:inc}).

The PDF of fully reconstructed events is similar to
$\P(\vec \xi)$, except that $\F_i(\fisher)$ is replaced by the
function $\E(\mes)$.
The fractions $f_{\DstAOne}^{\rm f}$, $f_{\peak}^{\rm f}$, and
$f_{{\rm \comb}\BB}^{\rm f}$ for this sample are determined from the
\mc\ simulation, and are of order a few percent. The fractions $f_{\rm
signal}^{\rm f}$ and $f_{\qqbar}^{\rm f}$ are obtained from a
3-dimensional fit to the $\mmiss$, $\mrho$, and $\mes$ distributions
of this sample.

The $\mmiss$ distribution of signal events is parameterized as a
bifurcated Gaussian, 
\begin{equation}
\M_{\rm signal}(\mmiss) \propto \exp 
	\left(-{(\mmiss -M)^2  \over  2 \sigma^2}\right),
\label{eq:bifur}
\end{equation}
where $M$ is the position of the peak, and the value of $\sigma$
depends on the sign of $\mmiss-M$. The proportionality constant in
this and subsequent PDF expressions is determined by integrating the PDF
over the allowed range of the PDF variable.
The $\mmiss$ distributions of the background event types are
parameterized as a bifurcated Gaussian plus an ARGUS
function~\cite{ref:argus},
\begin{eqnarray}
\A(\mmiss) &\propto& \mmiss \sqrt{1-\left({\mmiss /M_A}\right)^2} 
	\nonumber\\
	&\times& 
\exp\left[\epsilon \left(1-\left({\mmiss / M_A}\right)^2\right)\right],
\end{eqnarray}
where $\A(\mmiss) = 0$ for $\mmiss > M_A$, and $M_A$ and $\epsilon$
are parameters whose values are determined from fits to data or \mc\
simulation, as described in Sec.~\ref{sec:proc}.

The $\R_i(\mrho)$ functions are sums of a relativistic P-wave Breit
Wigner function and second-order polynomials.
The functions $\F_i(\fisher)$ are bifurcated Gaussians, and 
$\E_i(\mes)$ are a Gaussian for signal events and ARGUS
functions for the backgrounds.

The $\dt$ PDF of signal events is an exponential decay with the
$\Bz$ lifetime, convoluted with a triple-Gaussian resolution function
to account for finite detector resolution:
\begin{eqnarray}
\T_{\rm signal} (\Dt, \DtErr) &=&
{1 \over 2 \tauB} \int d\Dt_t\; 
	e^{-|\Dt_t| /\tauB} 
\times
	\biggl[f_n\, G_n(t_r, \DtErr)  \nonumber\\
&+& f_w\, G_w(t_r, \DtErr) 
	+ f_o\, G_o(t_r, \DtErr)\biggr],
\label{eq:sigDtPdf}
\end{eqnarray}
where $\Dt_t$ is the true decay time difference between the two $B$ mesons, 
$t_r \equiv \Dt - \Dt_t$ is the $\dt$ residual,
$G_n$, $G_w$, and $G_o$ are the ``narrow'', ``wide'', and
``outlier'' Gaussians, each of the form
\begin{equation}
G(t_r, \DtErr) \equiv 
	{1 \over \sqrt{2\pi} s \DtErr}
  \exp\left(-\,{\left(t_r - b\right)^2  
	\over 2 (s \DtErr)^2}\right),
\label{eq:Gaussians}
\end{equation}
where $s$ and $b$ are parameters obtained from the fit to data.
The coefficients $f_i$ of Eq.~(\ref{eq:sigDtPdf}) satisfy $f_w = 1 -
f_n - f_o$.  The same $\dt$ PDF parameters are used for signal and
$\aOne$ events.

The $\Dt$ PDFs of the combinatoric $\BB$, peaking $\BzBzb$, $\BDstst$,
and continuum are of the form
\begin{eqnarray}
\T_{\rm Bgd.}(\Dt, \DtErr) &=& 
(1 - f_o) \int d\Dt_t\; 
	\biggl[f_\tau {1 \over 2 \tau} \, e^{-|\Dt_t| /\tau}
		+ (1 - f_\tau)\, \delta(\Dt_t) \biggr]
	\nonumber\\ 
&\times &
	\biggl[f_n\, G_n(t_r, \DtErr) + 
	(1-f_n)\, G_w(t_r, \DtErr)\biggr] 
  + f_o 
	{1 \over \sqrt{2\pi} s_o }
  \exp\left(-\,{\left(t_r - b_o\right)^2  
	\over 2 s_o ^2}\right) ,
\label{eq:bgdDtPdf}
\end{eqnarray}
where the phenomenological parameter $\tau$ is
different from $\tauB$ of Eq.~(\ref{eq:sigDtPdf}). The $\delta(\Dt_t)$
term accounts for events in which the $\pi_h$ originates from
essentially the same point as the tracks dominating the determination
of $\ztag$. 
Parameter values for each of the different background PDFs are
determined from fits to independent control samples in data, as described in
Sec.~\ref{sec:proc}. 
The fraction $f_o$ of events corresponding to the outlier Gaussian is
taken to be zero for the three $\BB$ background PDFs, based on studies
performed with the \mc\ simulation and the data. In the continuum PDF,
$f_o$ is determined to be about 1\% for $s_o = 10$~ps.
In the peaking $\BzBzb$ PDF $f_\tau = 0$, reflecting the fact that
the $\pi_h$ originates from the decay of the other $B$.

\subsection{Fit Procedure}
\label{sec:proc}
The $\Bz$ lifetime $\tauB$ is obtained through a series of fits.
First, the kinematic variable PDF parameters of $\aOne$, combinatoric
$\BB$ and the peaking $\BzBzb$ background are determined by fitting
the distributions of these variables in the \mc\ simulated events. The
parameters of $\F_{\rm signal}(\fisher)$, as well as $f_{\DstAOne}$,
and $f_{\peak}$, are also obtained from the \mc\ simulation.

The $\R(\mrho)$, $\M(\mmiss)$, and $\E(\mes)$ parameters of signal and
continuum events, and the $\F(\fisher)$ parameters of continuum are
obtained by fitting the kinematic variable distributions of the data
in the signal region. This 4-dimensional fit is performed
simultaneously for partially and fully reconstructed on-resonance
data, and for off-resonance data.  The fractions $f_{\qqbar}$ and
$f_{{\rm \comb}\BB}$ are also determined in this fit, with $f_{\rm
signal}$ obtained from $\sum_i f_i = 1$.  The value of $f_{\Dstst}$ is
set to~0 in this and the subsequent fits, and is later varied when
studying systematic errors.

The parameters of $\T_{{\rm \comb}\BB}$ are determined by
fitting the data in the $\mmiss$ sideband. The
sideband is populated only by combinatoric $\BB$ and continuum
events. Consequently, this fit uses $\fisher$ as the only kinematic
variable, and is performed simultaneously for on- and off-resonance
data. The parameters of $\T_{\qqbar}(\Dt, \DtErr)$ of
this sample are also determined in the fit.

The composition of the wrong-sign event sample in the signal region
and the parameters of $\T_{\peak}(\dt, \DtErr)$ are determined from
fits to this sample. The parameters of $\T_{\qqbar}(\Dt, \DtErr)$ of
this sample are also determined in the fit.

Finally, the signal region parameters of $\T_{\rm signal}(\dt, \DtErr)$ and
$\T_{\qqbar}(\dt, \DtErr)$ are determined from a simultaneous fit to
the right-sign signal region on-resonance, off-resonance and fully
reconstructed data.  The parameters of $\T_{{\rm \comb}\BB}(\dt,
\DtErr)$ and $\T_{\peak}(\dt, \DtErr)$ are taken from the sideband and
wrong-sign fits, respectively.
Use of the sideband and wrong-sign samples for this purpose is
validated using the \mc\ simulation. 

The samples used to obtain parameters of the different $\dt$ PDF
components are summarized in Tab.~\ref{tab:proc}.

\begin{table}[!htb]
\vspace*{0.2truein}
\caption{Signal and background $\dt$ PDFs and the data samples from
which their parameters are determined. The indicated data samples are
used simultaneously in the fits, where ``on'' and ``off'' refers to on- and
off-resonance data, respectively, and ``full'' refers to the fully
reconstructed event sample.}
\label{tab:proc}
\begin{center}
\begin{tabular}{|l|l|l|l|}
\hline
PDF & Data samples & $\mmiss$ region & Right/wrong-sign \\
\hline
\hline
 $\T_{{\rm \comb}\BB}(\dt,\DtErr)$ &
 on $+$ off &
 Sideband  &
 Right-sign \\[0.05truein]
\hline
 $\T_{\peak}(\dt,\DtErr)$ &
 on $+$ off &
 Signal region &
 Wrong-sign \\[0.05truein]
\hline
 $\T_{\rm signal}(\dt, \DtErr)$   &
 on $+$ off $+$ full &
 Signal region &
 Right-sign \\[0.05truein]
$\T_{\qqbar}(\dt, \DtErr)$ & & & \\[0.05truein]
\hline
\end{tabular}
\end{center}
\end{table}

\section{Results}
The partially reconstructed signal region on-resonance sample contains
50898 events, including $5266\pm 251 \pm 34$ $\bdstrho$ and $691 \pm
36 \pm 4$ $\aOne$ events, as determined by the kinematic variable
fit. The systematic errors are due to the finite numbers of events in
the \mc\ simulation samples used to obtain the kinematic variable distributions
of the two $\BB$ backgrounds and $\aOne$. The fully reconstructed
sample contains $255 \pm 20$ signal events.

\begin{figure}[!htb]
\begin{center}
\includegraphics[width=16cm]{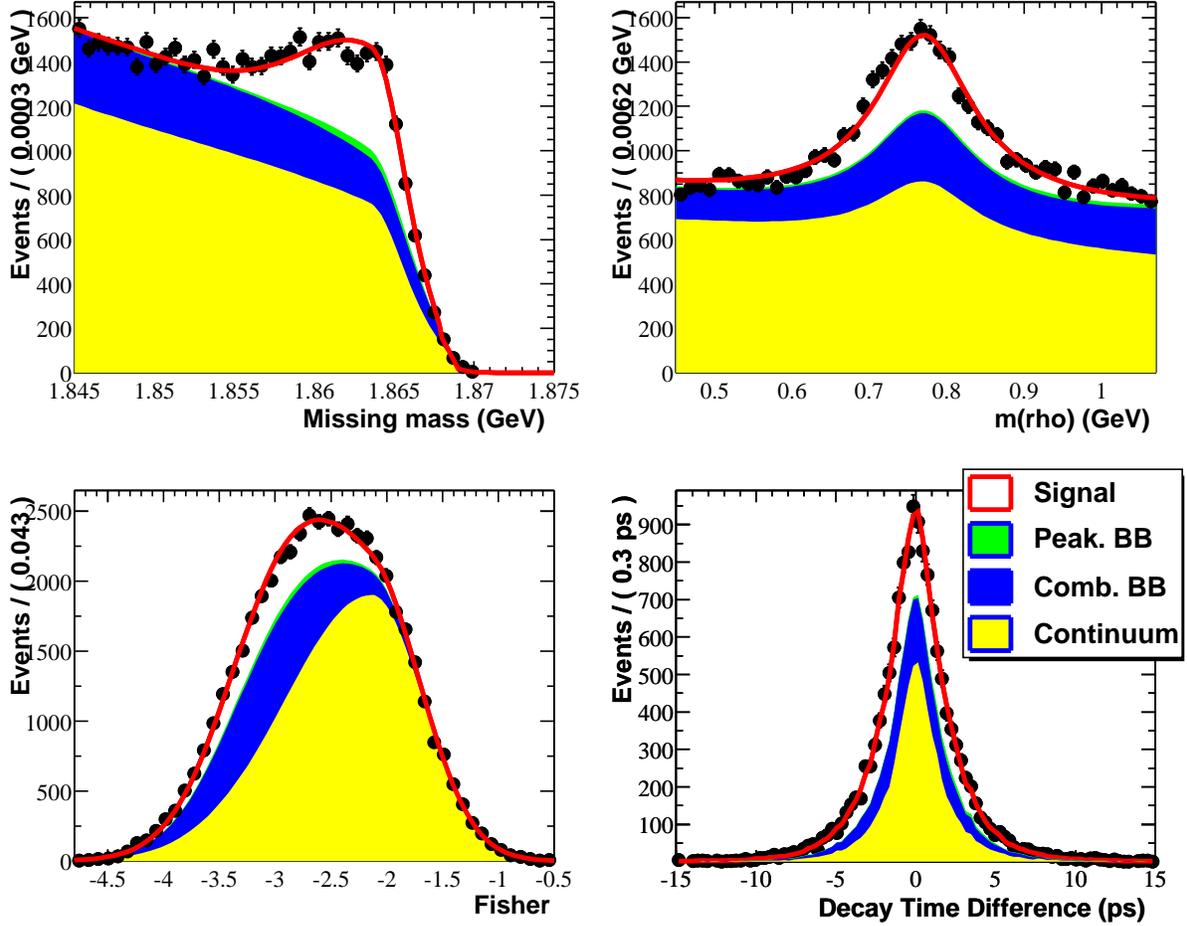}
\caption{Projections of the PDF of partially reconstructed events onto
	the on-resonance data in the variables $\mmiss$ (upper left), 
	$\mrho$ (upper right), $\fisher$ (lower left), 
	and $\dt$ (lower right). Events in the $\dt$ projection plot 
	satisfy the additional criteria $\mmiss > 1.854$~GeV, 
	$0.6 < \mrho < 0.93$~GeV, and $\fisher < -2.1$.}
\label{fig:results-on}
\end{center}
\end{figure}

\begin{figure}[!htb]
\begin{center}
\includegraphics[width=12cm]{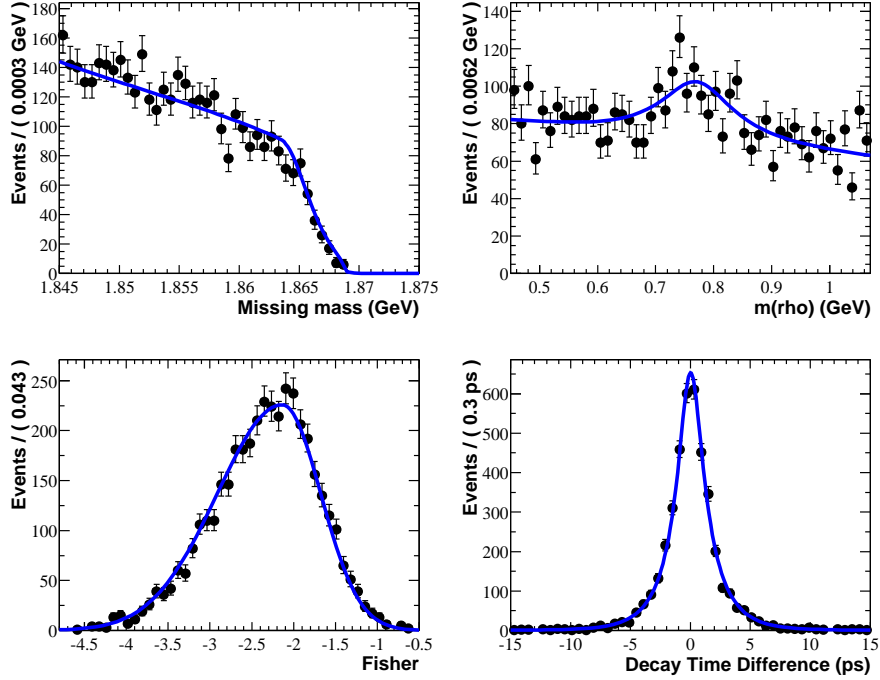}
\caption{Projections of the PDF of continuum events onto
	the off-resonance data in the variables $\mmiss$ (upper left), 
	$\mrho$ (upper right), $\fisher$ (lower left), 
	and $\dt$ (lower right).}
\label{fig:results-off}
\end{center}
\end{figure}

\begin{figure}[!htb]
\begin{center}
\includegraphics[width=12cm]{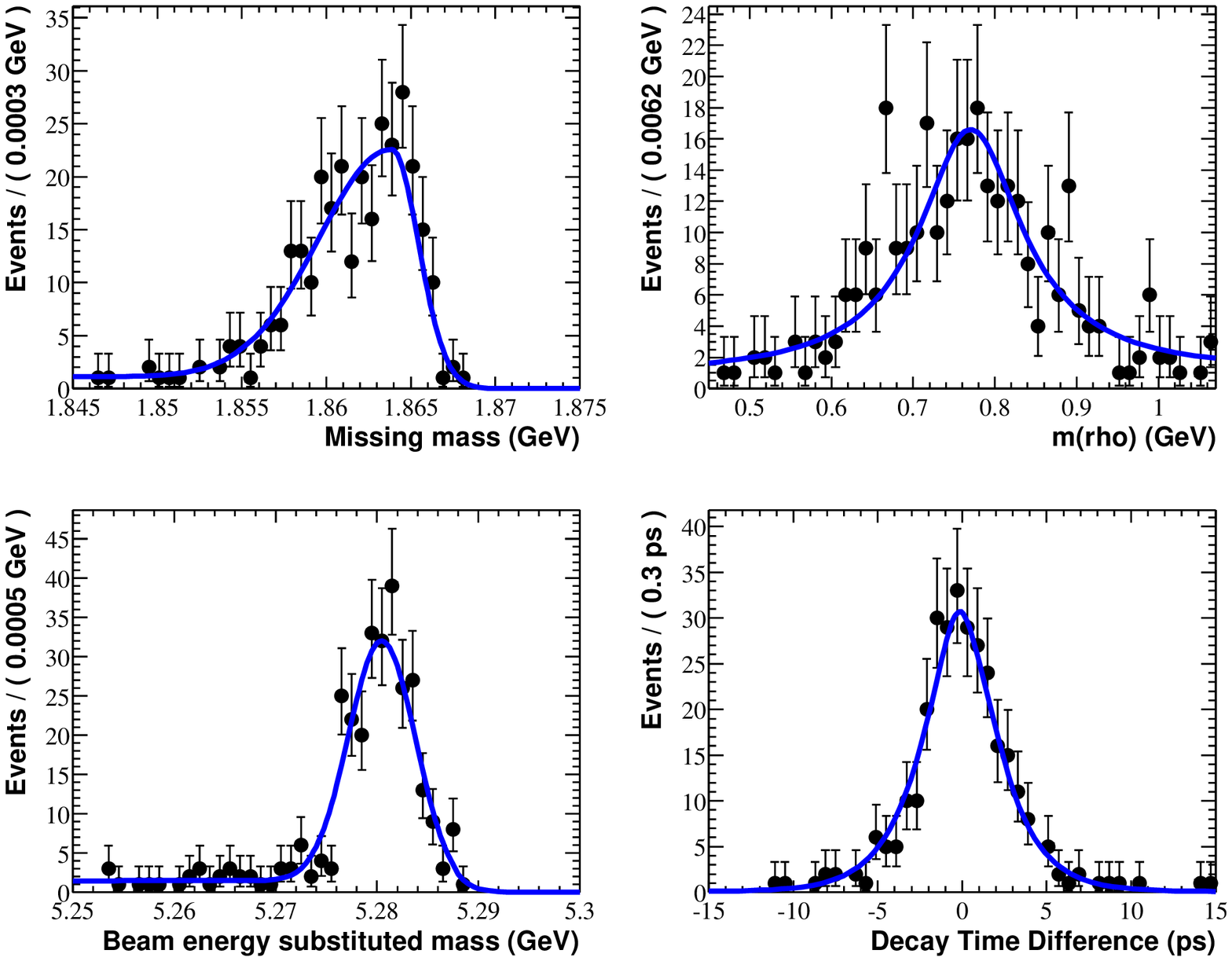}
\caption{Projections of the PDF of fully reconstructed events onto
	the fully reconstructed data  in the variables $\mmiss$ (upper left), 
	$\mrho$ (upper right), $\mes$ (lower left), 
	and $\dt$ (lower right).}
\label{fig:results-excl}
\end{center}
\end{figure}

Projections of the PDF onto the data are shown in Figs.~\ref{fig:results-on}
through~\ref{fig:results-excl}.
The result of the maximum likelihood $\dt$ fit is 
\begin{equation}
\tauB^{\rm raw} = 1.535 \pm 0.064~{\rm (stat.)}~ps,
\label{eq:tauraw}
\end{equation}
where the error is statistical only. 

Several corrections are applied to this result, in order to account
for known sources of bias. A correction of $+0.008 \pm 0.011$~ps is
due to the assignment of $a_1$ daughter tracks to the other $B$
vertex, decreasing the measured $\dt$ in $\aOne$ events. 
The resulting value of 1.543~ps is divided by $\RDz = 0.982 \pm
0.022$, the ratio between the $\Bz$ lifetime obtained from a fit to
signal \mc\ events and the value obtained from a fit to the true $\Dt_t$
distribution of this sample. This correction accounts for the effect
of $D$ daughter tracks that pass the cone cut and are used in the
other $B$ vertex.
A correction of $+0.014 \pm 0.013$~ps is applied to this result, to account
for a possible bias due to event selection, determined by fitting the 
true $\Dt_t$ distribution of signal events passing the selection criteria.
The magnitudes and errors of all these corrections are obtained from the
\mc\ simulation.

The fit PDF was used to generate and fit hundreds of \mc\ samples,
each corresponding to the data sample in number of events and PDF
parameters. The value of $\tauB$ obtained from these fits was on
average lower than the generated value by $0.031 \pm 0.005$~ps.
Repeating these studies with different \mc\ sample sizes, this bias is
understood to be due to limited sample statistics.  A correction of
this magnitude is therefore added to the value of $\tauB$. The fully
corrected result is
\begin{equation}
\tauB = 1.616 \pm 0.064~{\rm (stat.)~ps}.
\label{eq:tau-corrected}
\end{equation}

\section{Systematic Errors and Cross Checks}

\begin{table}[!htb]
\vspace*{0.2truein}
\caption{Systematic errors.}
\label{tab:syst}
\begin{center}
\begin{tabular}{|l|c|}
\hline
Source & Error (ps) \\
\hline
Statistical error of sideband fit & $ 0.033$    \\
Statistical error of kinematic fit & $ 0.029$   \\
Statistical error of wrong-sign fit & $ 0.002$  \\
\hline
\mc\ statistics: Calculation of $\RDz$ & $ 0.036$               \\
\mc\ statistics: Kinematic parameter fits & $ 0.014$  \\
\mc\ statistics: Event selection bias  & $ 0.013$ \\
\mc\ statistics: $\aOne$ bias  & $ 0.011$ \\
\hline
$\NDtr$ uncertainty & $ 0.026$ \\
Level of $\BDstst$ background & $0.023$  \\
Likelihood fit bias & 0.016 \\
Variation of fixed parameters & $ 0.015$ \\
${\cal B}(\aOne) / {\cal B}(\bdstrho)$  & $ 0.005$ \\
Level of peaking background & $ 0.003$ \\
Bias from fully reconstructed events & $ 0.001$ \\
\hline
SVT misalignment &  $ 0.008$ \\
$z$-length scale uncertainty & $ 0.007$ \\
Beam energies uncertainty&   $ 0.002$ \\
\hline
Total & $ 0.075$ \\
\hline
\end{tabular}
\end{center}
\end{table}

Several sources of systematic error are considered. 
The statistical error matrix obtained from the wrong-sign
signal region $\dt$ fit is used to vary the parameters of the peaking
$\BzBzb$ background, taking into account their correlations. A fit to the
right-sign signal region data follows each variation in these
background parameters. The resulting variations in $\tauB$ are
added in quadrature to form the total systematic error due to the
finite number of events in the wrong-sign sample.
With the same procedure, the errors due to the sideband fit are
propagated to the wrong-sign signal region and then to the right-sign
signal region fits, to obtain the error due to the sideband sample size.
The errors due to the finite number of events used in the kinematic variable
fits on data and the \mc\ simulation are evaluated in the same way.

The \mc\ statistical errors in the determination of $\RDz$, the
$\aOne$ bias, and the selection bias corrections are taken into account.
The fraction of events in which $D$ daughter tracks are used in the
other $B$ vertex fit is varied by $\pm 5\%$. The magnitude of this
variation is determined by comparing the $\NDtr$ distributions of fully
reconstructed data and \mc\ events. The resulting variation in $\RDz$
is used to evaluate the systematic error due to this uncertainty.

The contribution of the $\BDstst$ background is neglected in the
fits. To evaluate the systematic error associated with this, we
instead take the number of $\BDstst$ in the data sample to be 2400,
corresponding to 
${\cal B}(\BDstst) \; {\cal B}(\Dstst \rightarrow D^{*+} \pi) =
0.3\%$. This value is estimated from known branching fractions of the decays
$\overline B\rightarrow D^{**} \pi^-$, 
$\overline B\rightarrow D^{(*)} \rho^-$, 
$\overline B\rightarrow D^{(*)} \pi^-$, 
and available limits on 
${\cal B}(\overline B\rightarrow \Dstst \rho^-)$~\cite{ref:pdg}.
The PDF parameters of this background are taken from the \mc\ simulation. 
Repeating the $\dt$ fit yields a 0.023~ps change in the value of
$\tauB$, which is taken as the systematic error.
The $\Dstst$ states simulated are $D_1(2420)$, $D_2^*(2460)$, and
$D_1(j={1 \over 2})$, the latter having mass 2.461~GeV and width
290~MeV.  No significant difference is found between these states in
terms of their effect on the observable quantities of this analysis.

Two variations of the signal $\dt$ PDF are used in fits to the
data. In one variation, the parameter $b$ in Eq.~(\ref{eq:Gaussians})
is replaced by $b \DtErr$. In the other, the sum of the narrow and
wide Gaussians of Eq.~(\ref{eq:sigDtPdf}) is replaced by a Gaussian
convoluted with an exponential. The effect of generating \mc\ event
samples using one PDF and fitting them using another PDF was
studied. Based on these data and \mc\ studies, a systematic error of
$0.016$~ps is estimated due to the choice of signal $\Dt$ PDF.

The parameters $s$ and $b$ (Eq.~(\ref{eq:Gaussians})) of the signal
and continuum PDF outlier Gaussians are fixed in the fit to the
right-sign signal region data. To estimate the associated systematic
errors, their values are varied within reasonable ranges, and the
resulting changes in $\tauB$ are taken as systematic errors.

Additional systematic errors are due to the uncertainty in the
relative branching fractions of $\aOne$ and $\bdstrho$,
the level of peaking background, 
the introduction of a possible bias due to the use of fully
reconstructed events,
detector alignment and $z$-length calibration,
and beam energy uncertainty.
The total systematic error is 0.075~ps, dominated by the errors due to
sideband and kinematic fit statistical errors, and \mc\ statistical errors.
The systematic errors are listed in Tab.~\ref{tab:syst}.

Several cross-checks were conducted to ensure the validity of the
result. The number of signal events detected is in good agreement with
the published branching fraction~\cite{ref:pdg} and our signal
reconstruction efficiency.  The fit was repeated with different values
of the cone cut, ranging between 0.6 and 1.2~radians. The data were
fitted in bins of the lab frame polar angle, azimuthal angle, and
momentum of the $\pi_h$, and in sub-samples corresponding to different
SVT alignment calibrations. In all cases, no significant variation of
the result was observed.

\section{Conclusion}
In a sample of \nB\ $\BB$ pairs, we identified $5521\pm 252 \pm 34$
$\bdstrho$ and $691 \pm 36 \pm 4$ $\aOne$ events using partial and full $B$
reconstruction.  These events were used to measure the $\Bz$ lifetime,
with the preliminary result being
\begin{equation}
\tauB = 1.616 \pm 0.064~{\rm (stat.)} \pm 0.075~{\rm (syst.)~ps}.
\label{eq:tau-conclusion}
\end{equation}
This result is in good agreement with earlier published
measurements~\cite{ref:pdg}, constituting a necessary step in
validating the use of partially reconstructed $\bdstrho$ events for the
measurement of $\sin(2\beta + \gamma)$.

\bigskip
\bigskip
\parindent=0pt 
{\Large \bf Acknowledgements}
\bigskip

We are grateful for the 
extraordinary contributions of our \pep2\ colleagues in
achieving the excellent luminosity and machine conditions
that have made this work possible.
The success of this project also relies critically on the 
expertise and dedication of the computing organizations that 
support \babar.
The collaborating institutions wish to thank 
SLAC for its support and the kind hospitality extended to them. 
This work is supported by the
US Department of Energy
and National Science Foundation, the
Natural Sciences and Engineering Research Council (Canada),
Institute of High Energy Physics (China), the
Commissariat \`a l'Energie Atomique and
Institut National de Physique Nucl\'eaire et de Physique des Particules
(France), the
Bundesministerium f\"ur Bildung und Forschung
(Germany), the
Istituto Nazionale di Fisica Nucleare (Italy),
the Research Council of Norway, the
Ministry of Science and Technology of the Russian Federation, and the
Particle Physics and Astronomy Research Council (United Kingdom). 
Individuals have received support from 
the A. P. Sloan Foundation, 
the Research Corporation,
and the Alexander von Humboldt Foundation.

\end{document}